\documentclass[prl,twocolumn,groupedaddress,showpacs]{revtex4}
\usepackage{graphicx}
\usepackage{dcolumn}
\usepackage{amssymb,amsmath}

\begin{document}

\title{Sudden Collapse of a Granular Cluster}
\author{Devaraj van der Meer, Ko van der Weele, and Detlef Lohse}
\affiliation{Department of Applied Physics and J.M. Burgers Centre for
  Fluid Dynamics, University of Twente, P.O. Box 217, 7500 AE Enschede, The Netherlands}
%\date{\today}
%\pacs{45.70.-n, 02.30.Oz}

\sloppy

\begin{abstract}
Single clusters in a vibro-fluidized granular gas
in $N$ connected compartments become unstable at strong
shaking. They are experimentally shown to collapse \textit{very
  abruptly}. The observed cluster lifetime (as a function of the driving
intensity) is analytically calculated within a flux model, making use
of the self-similarity of the process. After collapse, the cluster diffuses
out into the uniform distribution in a self-similar way, with an
\textit{anomalous} diffusion exponent $1/3$.       
\end{abstract}

\maketitle

%\section{Introduction}

One of the key features of a granular gas, making it fundamentally
different from ordinary molecular gases, is its tendency to
spontaneously separate into dense and dilute regions
\cite{first6}. This clustering originates from the
dissipative nature of the particle collisions. It is an unwanted effect
in many applications where granular material is brought
into motion. Therefore we study (within a simple geometry) how
\textit{declustering} occurs. We find that the breakdown of a cluster
can be very abrupt, making declustering very different from clustering
in reverse time order.

The experimental system (see Fig. 1) consists of a row of $N$ equal
compartments, separated by walls of height $h=25.0$ mm and filled with
a few hundred steel beads of diameter 3.0 mm. We start out with all the
particles in the middle compartment and bring them into a gaseous
state by shaking the system vertically. For weak shaking the cluster
is stable: after some initial spilling, a dynamical equilibrium is
established between the outflux of slow particles from the cluster and
the influx of fast particles from outside
\cite{eggers99,vdweele01,vdmeer01}. For sufficiently strong shaking,
however, the cluster breaks down.  

\begin{figure}
\begin{center}
\includegraphics*[scale=1]{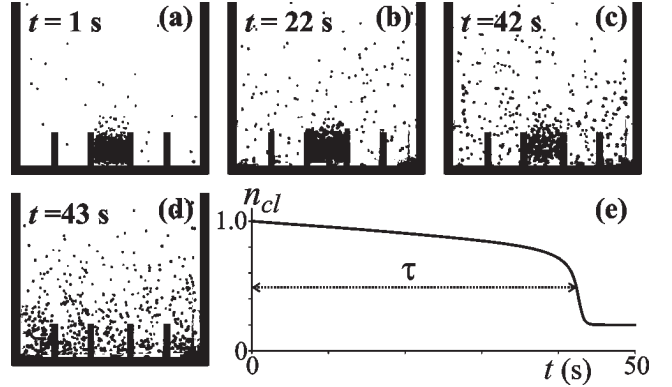}
%\begin{small}
\caption{\small Four images from a 5-box experiment, at
    driving parameter $\widetilde{B}=8.2$. The cluster is clearly present until
  $t=42$ s, then suddenly collapses, leaving no trace one second
  later. Plot (e) shows the time evolution of the cluster fraction
  $n_{cl}(t)$, evaluated from the flux model.}
%\end{small}
\label{fig-1}
\end{center}
\end{figure}

Two different
regimes are observed: (i) At very strong shaking the breakdown occurs
immediately, and the cluster spreads out over the boxes with its profile
widening as $t^{1/3}$ (instead of the standard $t^{1/2}$ diffusion
law). (ii) At less violent shaking, the cluster seems to remain stable
for a long time, showing only a small leakage to its neighbors. But
\textit{suddenly} it collapses and subsequently diffuses over all
boxes. The sudden death of the cluster is depicted in Fig.~\ref{fig-1}.

One thing this figure shows is that the breakdown of a cluster is
strikingly different from the reverse process of cluster formation,
which is known to take place gradually and (for all $N \geq 3$) via transient
states showing clusters in several boxes
\cite{vdweele01,vdmeer01}. This lack of time-reversibility is yet
another consequence of the dissipation in the system.   

\begin{figure}
\begin{center}
\includegraphics*[scale=1]{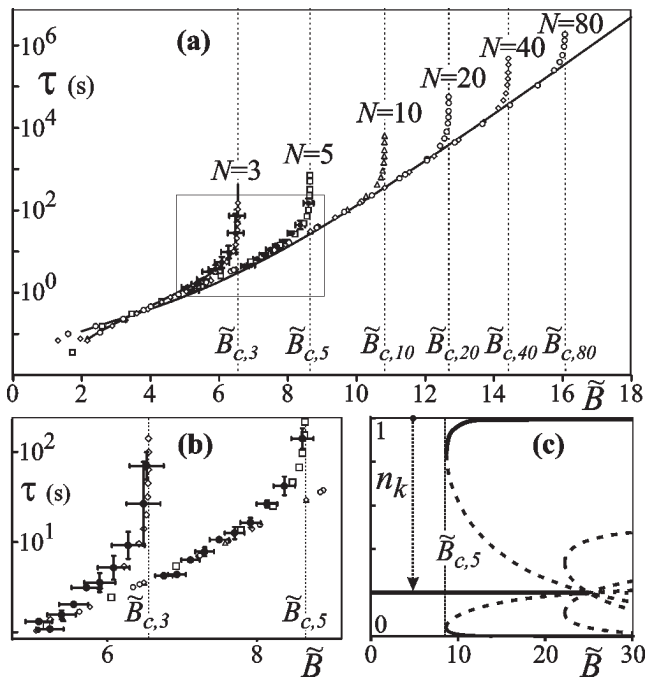}
\caption{\small   
(a) Lifetime $\tau$ vs driving parameter $\widetilde{B}$,
  determined from experiments with $N=3,5$ boxes
  (dots with error bars), and from numerical evaluation of the
  flux model with $N=3,5,10,20,40,80$ boxes (empty symbols). The solid curves are
  analytical solutions for $N=3$ \cite{art3note2} and for the envelope
  curve, which goes roughly as exp$((3/2)\widetilde{B})$
  (cf. Eq.~\ref{eq-tau}). (b) Enlargement of (a), showing the
  experimental results for $N=3$ and $N=5$ in more detail. Every point
  is based on 15 repetitions of the experiment; the vertical error bars denote
  the maximal deviation from the average $\tau$ measured, the horizontal ones
  represent the accuracy in $\widetilde{B}$.  
(c) Bifurcation diagram for $N=5$, showing
the critical value $\widetilde{B}_{c,5}$. Declustering occurs to the
left of this value.}
\label{fig-2}
\end{center}
\end{figure}

The abruptness of the collapse allows us to define a cluster lifetime
$\tau$ (via $\ddot{n}_{cl}(\tau)=0$,
see Fig.~\ref{fig-1}e). In Fig.~\ref{fig-2} the measured lifetimes are
plotted as a function of the inverse shaking strength $\widetilde{B}$
(Eq.~\ref{eq-wideB}) for various values of $N$. The data lie on a
universal envelope curve, until at some critical value
$\widetilde{B}_{c,N}$ (which grows with $N$)
they diverge.\\  

All the above experimental observations can be accounted for
\textit{quantitatively} by means of the flux model of
refs. \cite{eggers99,vdweele01,vdmeer01}.
At the heart of this model is a flux function $F(n_{k})$, describing the
outflow from the $k$th box to each of its neighbors. It is a non-monotonic
function of $n_k$ (the particle fraction in the box): $F(n_k)$ first
increases with $n_k$, but beyond a certain value of $n_k$ it decreases
again, as the increasingly frequent
\textit{inelastic} collisions slow the particles down, so that they
cannot make it over the wall to the neighboring compartments
anymore. The precise form of $F(n_k)$ is not very critical, as long as
it is a one-humped function. We will use Eggers' flux function \cite{eggers99}

\begin{equation}
\label{eq-grflux}
F(n_{k})=C \sqrt{\widetilde{B}}n_{k}^{2}e^{-\widetilde{B}n_{k}^{2}},  
\end{equation}
with
\begin{equation}
\label{eq-wideB}
\widetilde{B} \propto \frac{g h r^4 (1-e)^2 P^2}{S^2 (a f)^2} . 
\end{equation}

The driving parameter $\widetilde{B}$ depends on the total number of
particles $P$ and their properties (radius $r$, restitution 
coefficient $e$ of the particle collisions), on the geometry 
of the system (height $h$ of the walls, ground area $S$ of
each box), and on the frequency $f$ and amplitude $a$
of the shaking. The factor $C$ only determines the absolute rate of the
flux, and can be incorporated in the time scale. 

The equation of motion for the fraction in box $k$ is \cite{art3note1}  
\begin{equation}
\label{eq-netflux}
\frac{dn_{k}}{dt} = F(n_{k-1})-2F(n_{k})+F(n_{k+1}) ,
\end{equation}
\noindent where $k=1,2,..,N$. Here we assume a nearest neighbor
interaction, and a cyclic arrangement of the boxes ($k=N+1$ equals
$k=1$). We further impose particle conservation, $\sum_k n_k = 1$.

The numerical results shown in Fig.~\ref{fig-1}e and Fig.~\ref{fig-2}
have been obtained using the above flux model, starting out with all
particles in one box (labeled $cl$). They quantitatively agree with
the experimental observations. The decaying cluster goes through
\textit{three} different stages:
  
The starting stage is a very short one, in which both $n_{cl}$ and
$F(n_{cl})$ display a jump compared to $n_i$ and
$F(n_i)$ in the surrounding boxes, $i=1,2,..$ (we have to
consider one side only because of the symmetry in the system).
 
In the second stage, the flux has become continuous
but the particle fraction remains discontinuous. However, its
low-density counterpart $n_0$ (defined by $F(n_0) = F(n_{cl})$)
\textit{does} continuously connect to $n_1$. We will use this fact
later in the analysis of the envelope curve. The flux gradually grows,
and eventually $F(n_{cl})$ reaches its maximum value. This is
accompanied by rapid density changes and the sudden death of the
cluster at the lifetime $\tau$.

\begin{figure}
\begin{center}
\includegraphics*[scale=1]{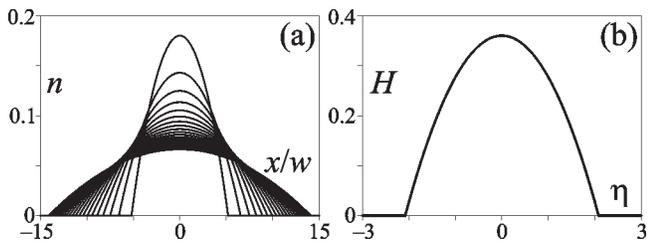}
\caption{\small (a) The diffusing profile at successive times $t$ (in the
  limit $\widetilde{B} = 0$). (b) The function $H(\eta)$ onto which all the
  curves in plot (a) collapse by a proper rescaling of the axes ($H
  \propto n t^{1/3}$ and $\eta
  \propto x t^{-1/3}$).} 
\label{fig-5}
\end{center}
\end{figure}

In the third and last stage, both $n_{cl}$
and $F(n_{cl})$ fit continuously to the other boxes (see
Fig.~\ref{fig-5}a). The remains of the cluster diffuse over the
whole system until the uniform distribution is reached.\\      

In what follows we will analytically solve the flux model. First we
focus on the third stage. We rewrite the problem into its continuum
version, by setting $n(x,t) \equiv n_k(t)$ ($x= k w$ by definition, where
the box width $w$ will be incorporated in the $x$-scale). Eq.~\ref{eq-netflux} then becomes: 
\begin{equation}
\label{eq-cont}
\begin{aligned}
\partial_t n &= \partial_{xx}F\big(n(x,t)\big)\\
&= C \sqrt{\widetilde{B}}
\partial_{xx}\left(n(x,t)^2e^{-\widetilde{B}n(x,t)^2} \right),
\end{aligned}
\end{equation}

\noindent and the conservation condition takes the form
$\int_{-\infty}^\infty n(x,t)dx = 1$.

For very strong shaking (regime (i), where $\tau$ is
vanishingly small) the diffusive stage sets in almost
immediately. Here $\widetilde{B} \to 0$, and Eq.~\ref{eq-cont} reduces
to
\begin{equation}
\label{eq-zeroB}
\partial_t n = C \sqrt{\widetilde{B}}\partial_{xx} (n^2) = 2C \sqrt{\widetilde{B}}((\partial_x n)^2+n \partial_{xx} n),
\end{equation}

\noindent which is known as the porous media equation \cite{barenblatt96,barwit}. The decay of the cluster in this limit is depicted in Fig.~\ref{fig-5}a.
It is self-similar: all curves in Fig.~\ref{fig-5}a fall onto a single
curve if we properly rescale the axes (Fig.~\ref{fig-5}b). The
original partial differential equation (PDE) can thus be brought back
to an ordinary differential equation (ODE) in terms of the
self-similarity variable $\eta = x/(C \widetilde{B}^{1/2} t)^{1/3}$.
With $n(x,t) = H(\eta)/(C \widetilde{B}^{1/2} t)^{1/3}$,
Eq.~\ref{eq-zeroB} now takes the form:
\begin{equation}
\label{eq-peaknon}
\begin{aligned}
\partial_{ \eta \eta}(H^2) + \frac{1}{3}\partial_{\eta}(\eta H) = 0.
\end{aligned}
\end{equation}
Its symmetric solution is $H(\eta)=H_0 -
(1/12)\eta^2$ (with the constant $H_0=(3^{1/3})/4 \approx 0.361$
determined by $\int_{-\infty}^{\infty} H(\eta) d\eta = 1$). This inverted parabola, depicted in Fig.~\ref{fig-5}b, represents in
one curve \textit{all} the stages of Fig.~\ref{fig-5}a. The scaling of the
axes shows that the height of the cluster decreases as
$t^{-1/3}$, and its width grows as $t^{1/3}$. This anomalous
diffusion (with exponent $1/3$) is also found in porous
  media \cite{barenblatt96,barwit}.
 The slowed down diffusion of the front originates from the quadratic
 $n$-dependence in $\partial_{xx} n^2$ (Eq.~\ref{eq-zeroB}). In the
 context of our granular model this reflects that particles only
 diffuse to neighboring boxes through 2-particle collisions: In the
 strong shaking limit the absence of particles slows down further diffusion, the presence enhances it.     
 
For non-zero $\widetilde{B}$ (regime (ii), where the diffusive stage has to
wait until after the sudden death) we have an additional dimensionless
variable $\chi = \widetilde{B}/C (t-\tau)$. Its
influence diminishes with time and the solutions converge to
the inverted parabola of the case $\widetilde{B} \to 0$.\\

Next we turn to the second, semi-continuous stage. For moderate
shaking, $\widetilde{B} \approx \widetilde{B}_{c,N}$, this stage can
take quite a long time. At the critical point $\widetilde{B}_{c,N}$
the lifetime $\tau$ even diverges to infinity and  the cluster
becomes stable. Starting out from the initial {..,0,0,1,0,0,..}
distribution, the system first \textit{very slowly} approaches a
distribution in which $n_{cl}$ is close to the cluster density at the
saddle-node bifurcation (see Fig.~\ref{fig-2}c) and 
all other boxes contain equal fractions $n_k=(1-n_{cl})/(N-1)$. Only when it has passed this
phantom equilibrium (i.e., when $n_{cl}$ is below the level of the
saddle-node bifurcation), the system quickens its pace and the sudden
collapse occurs.

This means that $\tau$ is the time it takes to pass the phantom
equilibrium. It can be calculated either numerically or analytically
(by integrating the Taylor expansion of
Eq.~\ref{eq-netflux}), with the result $\tau \propto
(\widetilde{B}_{c,N} - \widetilde{B})^{-1/2}$. So $\tau$ diverges as
the inverse square root of the distance to the critical point, which is the
common (mean field) power-law behavior near a second order phase
transition as we have here for $\tau$ \cite{landau80}.\\

Finally, we calculate the lifetime away from the critical
point, i.e., for $\widetilde{B}$ considerably smaller than $\widetilde{B}_{c,N}$. This will give us an analytical
expression for the envelope curve in Fig.~\ref{fig-2}a. For these
$\widetilde{B}$ values, the collapse occurs before the particles
leaking out of the cluster have had time to fill the outermost boxes
to any significant level. Therefore, the
behavior does not depend on the value of $N$: The system does not feel
its finite size during the cluster's lifetime, so the number of boxes
can be taken to be infinite.

The time-evolution of the cluster is described by
Eq.~\ref{eq-netflux}, with $F(n_{-1})=F(n_1)$:  
\begin{equation}
\label{eq-between}
\frac{dn_{cl}}{dt} = -2 F(n_{cl}) + 2 F(n_1) = -2 F(n_0) + 2 F(n_1).
\end{equation}

\noindent This equation contains $n_1$, which is governed by a similar equation of motion
(Eq.~\ref{eq-netflux}) containing $n_2$, etc. So we have to
deal with an infinite number of coupled nonlinear ODE's \cite{art3note2}.
This is a problem that cannot be solved directly, so we attack it in
five steps.

\textit{Step 1}: We first rewrite the problem into its continuum
version, and replace the cluster density $n_{cl}$ by its low-density
counterpart $n(0,t) \equiv n_0(t)
\approx n_{cl}$exp$(-\widetilde{B}n_{cl}^2/2)$. Thus, without
influencing the fluxes (since $F(n_0) = F(n_{cl})$) we make
$n(x,t)$ continuous in $x=0$. The density $n(x,t)$ obeys
Eq.~\ref{eq-cont}, plus a conservation condition saying that the
increase of material into the rest of the system equals the
influx from $x=0$: 
\begin{equation}
\label{eq-cont2}
\begin{aligned}
% \partial_t n &= w^2
% \partial_{xx}F\big(n(x,t)\big),\\
% = w^2 C \sqrt{\widetilde{B}}
% \partial_{xx}\left(n(x,t)^2e^{-\widetilde{B}n(x,t)^2} \right),
\partial_t \int_0^\infty
n(x,t)dx &= -(\partial_x\big(F\big(n(x,t)\big)\big)_{x=0}.
\end{aligned}
\end{equation}

\textit{Step 2}: The cluster-equation Eq.~\ref{eq-between} now becomes:
\begin{equation}
\label{eq-between2}
\begin{aligned}
\frac{dn_{cl}}{dt} &= 2 \Big( \partial_x F\big(n(x,t)\big) \Big)_{x=0}\\
&= 2 F'\big(n(0,t)\big) \big( \partial_x n(x,t) \big)_{x=0}.
\end{aligned}
\end{equation}
\noindent Since $F'$ can be derived directly from
Eq.~\ref{eq-grflux}, the problem reduces to determining
$\partial_x n(x,t)$ at $x=0$. 

\begin{figure}
\begin{center}
\includegraphics*[scale=1]{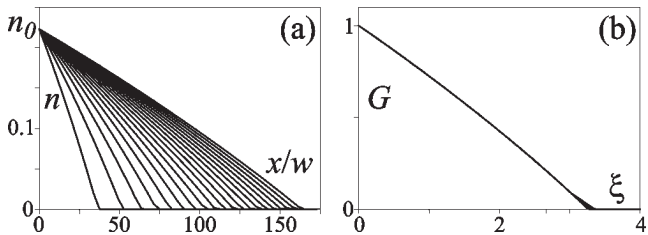}
\caption{\small (a) The density profile in the boxes surrounding the
  cluster, at 20 consecutive (equidistant) moments in time, within the
  constant $n_0$ approximation; $n_0=0.223$ and
  $\widetilde{B}=3.00$. (b) The function $G=n/n_0$ onto
  which all curves in plot (a) collapse by a proper rescaling of the
  axes ($\xi \propto x t^{-1/2}$.)} 
\label{fig-4}
\end{center}
\end{figure}

\textit{Step 3}: In order to do so, we observe that changes in
$n(0,t)$ happen on a much longer timescale than in the surrounding
boxes \cite{art3note3}, so the cluster acts
as a constant reservoir spilling granular material. This approximation
is illustrated in Fig.~\ref{fig-4}a: the profile in the system builds up
while $n_0$ remains constant. In fact, this build-up takes place in a
self-similar way (see Fig.~\ref{fig-4}b). So, as before, the problem
for $n(x,t)$ can be formulated
in terms of one variable $\xi = x/(n_0 C \widetilde{B}^{1/2} t)^{1/2}$.
Setting $n(x,t) = n_0 G(\xi)$, Eq.~\ref{eq-cont} becomes an ODE for
$G(\xi)$, and also the accompanying conservation condition
(Eq.~\ref{eq-cont2}) contains $\xi$ only: 
\begin{equation}
\label{eq-jetnon}
\begin{aligned}
\frac{1}{2}\xi \partial_{\xi}G &= - \partial_{\xi\xi}(G^2 e^{-\beta
  G^2}),\\
\left( \partial_{\xi}G \right)_{\xi=0} &= -\frac{e^{\beta}}{4(1-\beta)}
\int_{0}^{\infty} G(\xi) d\xi ,
\end{aligned}
\end{equation}     
\noindent where $\beta = \widetilde{B} n_0^2$.

\textit{Step 4}: The slope $(\partial_x n(x,t))_{x=0}$ can now be
approximated by $n_0 (t) (\partial_{\xi}G(\xi))_{\xi=0} (\partial_x
\xi)_{x=0}$, where we have revived the (slow) time-dependence in
$n_0(t)$. With $\partial_x \xi = (n_0(t) C \widetilde{B}^{1/2}
t)^{-1/2}$, Eq.~\ref{eq-between2} becomes:    
\begin{equation}
\label{eq-between3}
\frac{dn_{cl}}{dt} = 2\frac{\sqrt{n_0(t)}}{\sqrt{C \widetilde{B}^{1/2}t}} F'\big(n_0(t)\big) \big(
  \partial_{\xi}G(\xi)\big)_{\xi=0} .
\end{equation}
All quantities on the right hand side are tractable. We re-express
$n_0(t)$ in terms of $n_{cl}(t)$, derive $F'$ from
Eq.~\ref{eq-grflux}, and solve Eq.~\ref{eq-jetnon} to determine
$\big(\partial_{\xi}G(\xi)\big)_{\xi=0}$. This last step still
requires some work, because Eq.~\ref{eq-jetnon} does not allow an
analytical solution and moreover contains
$\beta$ (and hence $n_0(t)$) explicitly. Since $\beta$ is small,
however, we may expand
$G(\xi)$ and Eq.~\ref{eq-jetnon} in terms of $\beta$, and solve
numerically. In leading order we find $\big(\partial_{\xi}G(\xi)\big)_{\xi=0}=-K=-0.3138$. Inserting all this,
Eq.~\ref{eq-between3} becomes:
\begin{equation}
\label{eq-final}
%\begin{aligned}
\frac{dn_{cl}}{dt} =
- 4 \frac{K n_{cl}^{3/2}e^{-(3/4)\widetilde{B}n_{cl}^2}}{\sqrt{C \widetilde{B}^{1/2}t}}
\Big(
1-\widetilde{B}n_{cl}^2e^{-\widetilde{B}n_{cl}^2} \Big) . 
%\end{aligned}
\end{equation}

\noindent This is an ODE for $n_{cl}$ in closed form, which replaces
the original problem (Eq.~\ref{eq-between}) consisting of an infinite number of coupled ODE's.

\textit{Step 5}: Finally, we integrate Eq.~\ref{eq-final}
over the cluster density (dropping the suffix $cl$) and find an
analytical expression for the lifetime $\tau$ away from the critical points:  
\begin{equation}
\label{eq-tau}
\begin{aligned}
\tau = \Bigg[ \int_{n_{thr}}^1 \frac{\sqrt{C \widetilde{B}^{1/2}} n^{-3/2} e^{(3/4)\widetilde{B}n^2}dn}
{ 8 K \big( 1-\widetilde{B}n^2 e^{-\widetilde{B}n^2} \big)} 
\Bigg]^2 .
\end{aligned}
\end{equation}

\noindent Here $n_{thr}$ is the value of $n$ at which the
sudden death occurs. For the evaluation of the solid curve in
Fig.~\ref{fig-2}a we used $n_{thr}=0.5$, but this value is not too
critical (cf. Fig.~\ref{fig-1}e). The only free parameter is the
constant $C$: if this is adjusted properly, the analytical $\tau$ curve
agrees with the measured data over the whole range of $\widetilde{B}$-values.  

The above expression shows that $\tau$ roughly increases as
exp$(\widetilde{B}^{3/2})$. Recalling that $\widetilde{B}$ is the
inverse shaking strength, this underlines the experimental observation
that even a small reduction in the shaking strength causes a tremendous
increase of the cluster lifetime.\\

%In conclusion, in the studied compartmentalized system clusters break
%down very abruptly, in contrast to their slow formation. As clustering
%itself, the lack of time reversibility originates from the dissipative
%nature of the particle collisions. The dynamics is \textit
%{quantitatively} described by a remarkably simple flux model, which can be ana%lytically solved. 

In conclusion, in the studied compartmentalized system clusters break
down very abruptly, in contrast to their slow formation. As clustering
itself, the lack of time reversibility originates from the dissipative
nature of the particle collisions: The breakdown of the unstable
cluster is delayed because most of the energy input is dissipated
through collisions in the cluster. The dynamics is \textit
{quantitatively} described by a remarkably simple flux model, which can be analytically solved.   
\\

\begin{acknowledgements}      
\textit{Acknowledgments}: We would like to thank Ernst van Nierop for measuring the experimental
lifetimes. This work is part of the research program of the Stichting FOM, which is financially supported by NWO.
\end{acknowledgements}

\begin{small}

\end{small}


\begin{thebibliography}{10}
\bibitem{art3note1}
Statistical fluctuations, which would add a noise term to
Eq.~(\ref{eq-netflux}), are left out of this (mean field) description.
\bibitem{art3note3}
A cluster change $\Delta n_{cl}$ (flowing into the surrounding boxes)
corresponds to a much smaller change $\Delta n_0$ in its
low-density counterpart: $dn_{cl}/dn_0  = (1-\widetilde{B}n_{cl}^2)^{-1}
e^{\widetilde{B}n_{cl}^2/2} \approx
(1-\widetilde{B})^{-1}e^{\widetilde{B}/2}$ since $n_{cl} \approx
1$. So for most $\widetilde{B}$-values of interest (excluding the
strong-shaking regime $\widetilde{B} \to 0$) $|\Delta n_{cl}| =  \sum_{i
  \not = 0} \Delta n_i \gg \Delta n_0$.
\end{thebibliography}
\end{document}